\documentclass[journal=aaemcq,manuscript=article]{achemso}
\usepackage{graphicx}
\usepackage{bm}
\usepackage{mathptmx}
\usepackage{epstopdf}
\usepackage{booktabs}
\usepackage{multirow}
\usepackage{hhline}
\usepackage{dcolumn}
\usepackage[version=3]{mhchem}
\usepackage{comment}
\usepackage{textalpha}
\DeclareUnicodeCharacter{2212}{-}
\DeclareUnicodeCharacter{2009}{\,} 
\usepackage{subfig}
\usepackage{amsmath}
\usepackage{float}
\usepackage{chemformula}
\usepackage[version=3]{mhchem}
\usepackage{chemfig}
\usepackage{soul}

\makeatletter
\newcommand*{\forcekeywords}{
  \acs@keywords@print
  \let\acs@keywords@print\relax
}
\makeatother
\title{Controlling Moisture for Enhanced Ozone Decomposition: A Study of Water Effects on CeO$_2$ Surfaces and Catalytic Activity}

\author{Suchitra Gupta}
\affiliation{Indo-Korea Science and Technology Center (IKST),  Bangalore-560064, India}
\affiliation{Functional Ceramics Department
Korea Institute of Materials Science (KIMS)
797 Changwon-daero, Seongsan-gu, Chanwon-si, Gyeongsangnam-do, 51508, South Korea}
\author{Joon Hwan Choi}
\author{Hojin Jeong}
\affiliation{Functional Ceramics Department
Korea Institute of Materials Science (KIMS)
797 Changwon-daero, Seongsan-gu, Chanwon-si, Gyeongsangnam-do, 51508, South Korea}
\email{hjeong@kims.re.kr}
\author{Seung-Cheol Lee}
\affiliation{Indo-Korea Science and Technology Center (IKST),  Bangalore-560064, India}
\email{leesc@kist.re.kr}
\author{Satadeep Bhattacharjee}
\affiliation{Indo-Korea Science and Technology Center (IKST),  Bangalore-560064, India}
\email{s.bhattacharjee@ikst.res.in}
\keywords{Ozone decomposition, water poisoning, chemisorption, first-principles calculations}
\begin{document}
\begin{abstract}
This study investigates the catalytic degradation of ground-level ozone on low-index stoichiometric and reduced CeO$_2$ surfaces using first-principles calculations. The presence of oxygen vacancies on the surface enhances the interaction between ozone and catalyst by serving as active sites for adsorption and decomposition. Our results suggest that the \{111\} surface has superior ozone decomposition performance due to unstable oxygen species resulting from reaction with catalysts. However, when water is present, it competes with ozone molecules for these active sites, resulting in reduced catalytic activity or water poisoning. A possible solution could be heat treatment that reduces the vacancy concentration, thereby increasing the available adsorption sites for ozone molecules while minimizing competitive adsorption by water molecules. These results suggest that controlling moisture content during operation is crucial for the efficient use of CeO$_2$-based catalysts in industrial applications to reduce ground-level ozone pollution.
\end{abstract}

\section{Introduction}
Ozone exhibits antibacterial and antiviral properties and has a redox potential of 2.08 eV, making it a powerful oxidizing agent. These properties pave the way for applications such as water purification, pollutant degradation and inhibition of nitric acid production by fully oxidizing exhaust gases \cite{Hooman.2019, Camel.1998, kim2019grasping,Oh.2016}. This leads to the release of residual ozone in the atmosphere at ground level, which can negatively impact human health and damage crops. \cite{Camel.1998,Felzer.2007,Charles.2000,Sillman.1999,Solomon.2008, Post.2005,Turner.2016}. In the interests of human and environmental safety, reducing ozone pollution in industrial processes and in the atmosphere is of the utmost importance.
Activated carbon adsorption, chemical absorption, and catalytic decomposition are the existing
processes to remove the ground level ozone \cite{Gibbs.1998, Thomas.2011, Gong.2017, Chen.2022}. Among which catalytic decomposition of ozone into O$_2$ at low temperatures is the promising method of ozone elimination as 
it gives better efficiency and is also environment friendly process. Transition metal oxides and
noble metal catalysts are the commonly used catalysts for ozone decomposition. Owing to the high
cost of noble metals, transition metal oxides have attracted greater interest in the past. It has been
found that \textit{p}-type oxides like MnO$_2$, NiO, Co$_3$O$_4$, Fe$_2$O$_3$, Ag$_2$O, and CeO$_2$ 
shows greater catalytic efficiency as compared to the \textit{n}-type oxides which include 
V$_2$O$_5$, CuO, MoO$_3$ etc. \cite{Oyama.1997}.

Ozone can be adsorbed on metal oxide surfaces and dissociate into the reactive oxygen species, however, the oxide surfaces have a high affinity for water molecules \cite{Charles.2011,John.2005}. The oxide surface is hydroxylated by the chemical adsorption and dissociation of water. Depending on the surface structure, isolated hydroxyl groups, hydrogen-bonded hydroxyl groups, and bridged hydroxyl groups can be formed on different oxides \cite{Noei.2008, Larish.2014}. This can greatly affect the catalytic decomposition of ozone on oxides. Previous studies claim that surface hydroxyl groups are the site of ozone depletion, however, not all hydroxyl groups are catalytically active \cite{Zhang.2008,Ma.2009}. The presence of water has a major impact on the catalyst efficiency for ozone decomposition in transition metal oxides \cite{John.2005, Wang.2014, Liu.2019, Yan.2019}. 

In this work, using first-principles calculation we have studied the interaction of ozone on \{111\},
\{110\}, and \{100\} surfaces of CeO$_2$. Due to differences in the atomic arrangement, these surfaces
exhibit different catalytic performances. As oxygen vacancies are the adsorption sites for ozone
decomposition and their presence has also improved the rate of ozone elimination in oxides. In the
presence of oxygen vacancies, ozone behavior on the considered surfaces was also analyzed. It has
already been pointed out that the presence of water interferes with the catalytic activity of the catalyst, we
have looked into the effect of water on ozone adsorption on stoichiometric and oxygen-deficient
CeO$_2$\{111\} surface. We have also shown the effect of oxygen vacancies concentration on the
adsorption of ozone in the presence of water. Furthermore, we investigate the competitive co-adsorption of H$_2$O and O$_3$ on CeO$_2$ surfaces, examining their binding properties and potential interference with ozone decomposition.
We find that heat treatment may help optimize the performance of CeO$_2$-based catalysts in the dissociation of ground-level ozone. Heat treatment can reduce the number of oxygen vacancies on CeO$_2$ surfaces, making the surface more similar to a stoichiometric surface with no defects. However, as H$_2$O has larger adsorption energy on reduced \{111\} surfaces compared to its adsorption energies on stoichiometric \{111\}, \{110\}, and \{100\} surfaces, therefore, by reducing oxygen vacancies, the surface becomes not only more stable, it also becomes less reactive towards H$_2$O, which can improve the selectivity of ozone dissociation in the presence of water. 

\section{Computational Details}
Our first-principles calculations are based on density functional
theory as implemented in the Vienna Ab initio Simulation Package (VASP) \cite{Kresse.1994}. The exchange-
correlation energy is implemented by the Perdew-Burke-Ernzerhof (PBE) \cite{Perdew.1996}, and the core-valence
electron interaction is described using the projected augmented wave (PAW) method \cite{P.E.1994}. For Ce \textit{f} orbitals we have chosen U value of 5 eV throughout the calculations in consistent with the values
recommended in previous theoretical studies \cite{Nolan.2005,Sonja.2005, Zhou.2019,Marco.2012}. The \{111\}, \{110\}, and \{100\} slabs are
constructed using ($1\times1$), ($2\times1$), and ($2\times2$) conventional unit cells consisting of 12 (16 CeO$_2$ units),
4 (16 CeO$_2$ units), and 7 atomic layers (24 CeO$_2$ units) respectively. A vacuum gap of at least 12 Å
is included along the z-direction to minimize the interaction between periodic images. The energy
cut-off of 520 eV is used to truncate the plane wave basis sets and the entire Brillouin zone is
sampled using $4\times4\times1$ k-points mesh. The spin-polarized calculations are carried out for a reduced
system consisting of oxygen vacancies. The bottom layers are kept fixed, and the uppermost layers
are allowed to relax until the maximum forces on each atom are less than 0.03 eV/Å. The dipole
correction is applied along the z-direction to treat the artificial electric field arising due to the
asymmetry (resulting due to adsorbate adsorption and introduction of oxygen vacancies) and
periodicity of the slab.
\section{Results and Discussion}
CeO$_2$ crystallizes in the cubic fluorite structure with \textit{Fm3m} symmetry, and the slabs are modelled
using the lattice constant of 5.49 Å obtained theoretically which is in close agreement with the
experimental value of 5.41 Å \cite{Sims.1976} and previous theoretical results \cite{Sonja.2005,Loschen.2007,Shi.2016,Baudin.2004}. We have investigated
the \{111\}, \{110\}, and \{100\} surfaces of CeO$_2$, among which \{100\} surface is type III according
to Tasker classification i.e. it is polar \cite{Tasker.1979}. For the simulation of \{100\} surface, we have moved the half of surface
oxygen atoms to the other side of the slab to quench the dipole moment normal to the surface. CeO$_2$\{111\}
surface is most stable of all followed by \{110\}, and \{100\} surfaces \cite{Parker.1994,Jan.1998}. O atoms of ozone, water,
and surfaces are represented by $O_{ozone}$, $O_{water}$, and O respectively.
\par
In the following sections, we explore the adsorption and decomposition of ozone (O$_3$) on CeO$_2$ surfaces and the influence of surface structure and moisture on the catalytic activity of CeO$_2$-based catalysts. We find that O$_3$ adsorbed in a dissociative form is more thermodynamically favorable on the \{111\} surface compared to \{110\} and \{100\} surfaces. Oxygen vacancies on the reduced surfaces of CeO$_2$ serve as active sites for O$_3$ adsorption and decomposition, facilitating the process more efficiently than on stoichiometric surfaces. Additionally, the presence of water interferes with O$_3$ adsorption on reduced \{111\} surfaces by blocking the oxygen-deficient site, which is crucial for ozone adsorption and decomposition. These findings contribute to our understanding of ozone decomposition on CeO$_2$ catalysts and offer insights for optimizing their performance in various industrial applications.
\subsection{O$_3$ adsorption on stoichiometric \{111\}, \{110\}, and \{100\} surfaces}
We have considered the adsorption of O$_3$ on the stoichiometric CeO$_2$ surfaces and calculated
the adsorption energy (E$_{ads.}$) using the following relation:  
\begin{equation} 
E_{ads.} = E_{slab+molecule}-E_{slab}-E_{molecule}, 
\end{equation}
where, E$_{slab+molecule}$ is the energy of slab with molecule adsorbed on it, E$_{slab}$ and E$_{molecule}$ are the
energies of clean slab and molecule in the gaseous phase respectively. 

\begin{figure}[ht]
\centering
        \hspace{-1.8 cm}
	\subfloat[a]{\includegraphics[width= 0.26\textwidth]{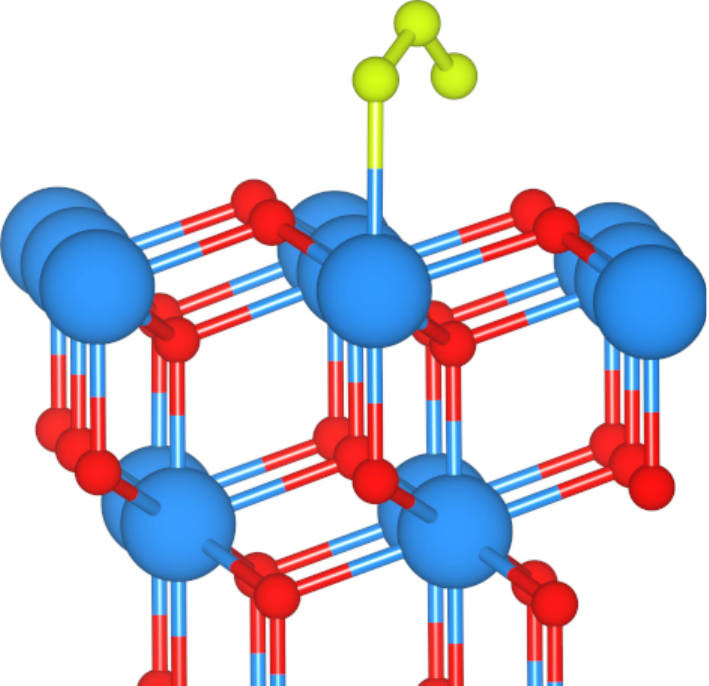}\label{fig: Figure 1a}}
         \hspace{0.6 cm}
         \subfloat[b]{\includegraphics[width= 0.34\textwidth]{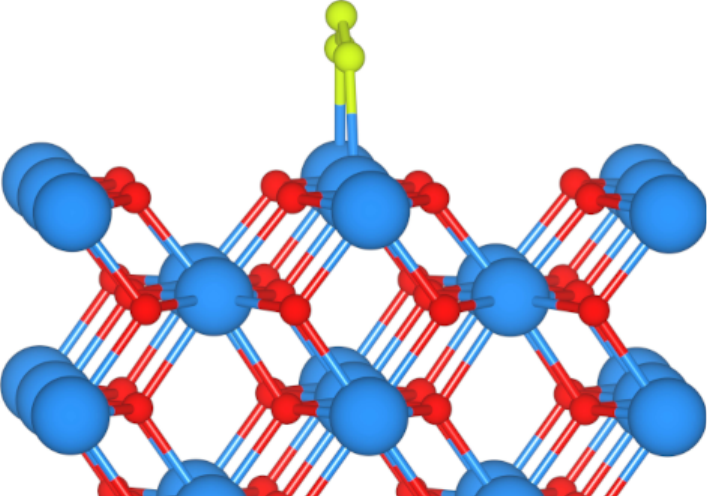}\label{fig: Figure 1b}}
	 \hspace{0.6 cm}
\subfloat[c]{\includegraphics[width= 0.37\textwidth]{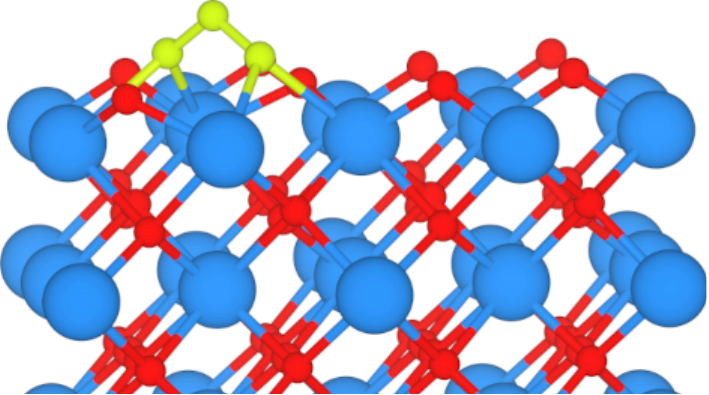}\label{fig: Figure 1c}}\\
	\hspace{-1.8 cm}
\subfloat[d]{\includegraphics[width= 0.31\textwidth]{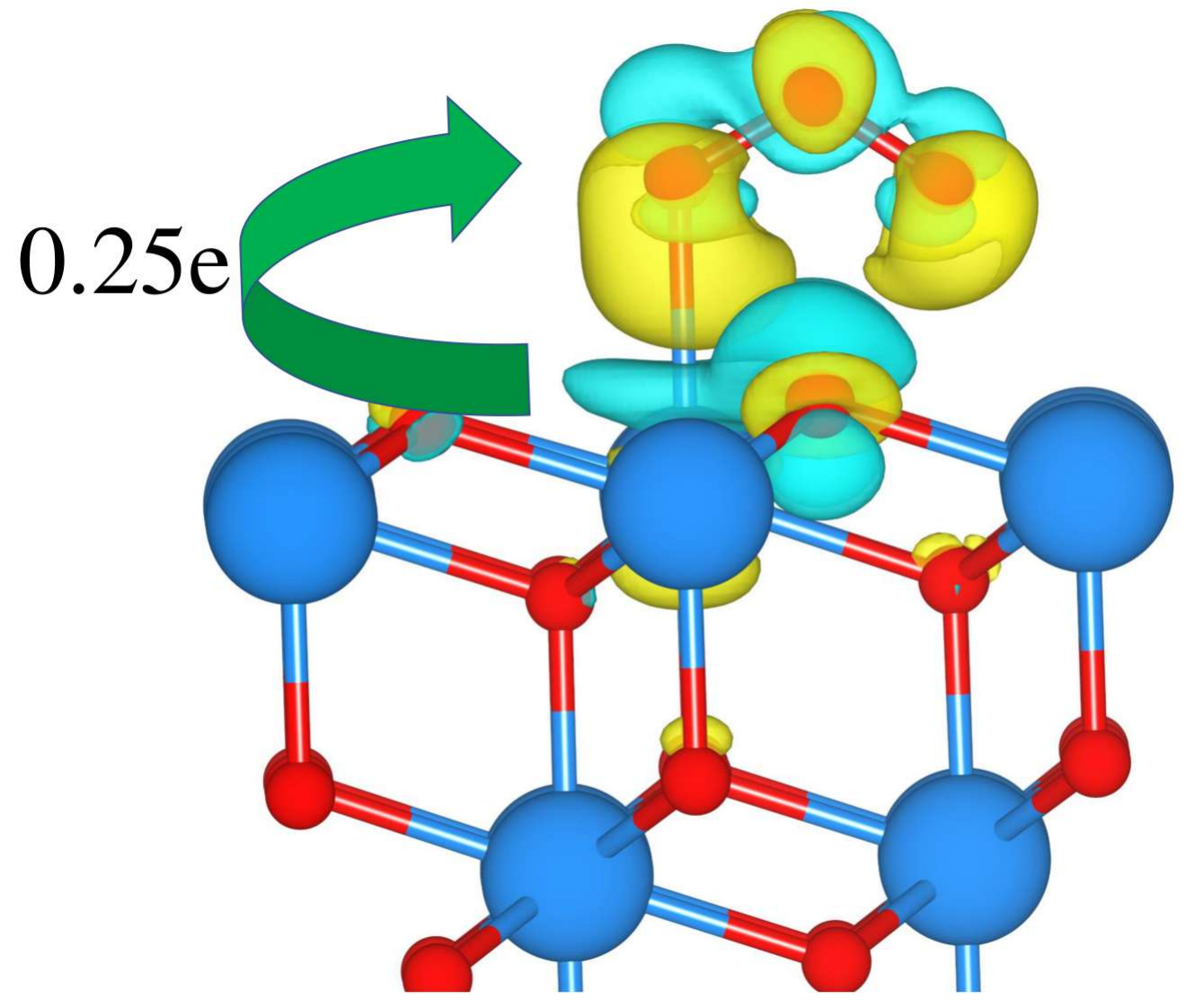}\label{fig: Figure 1d}}
\subfloat[e]{\includegraphics[width= 0.35\textwidth]{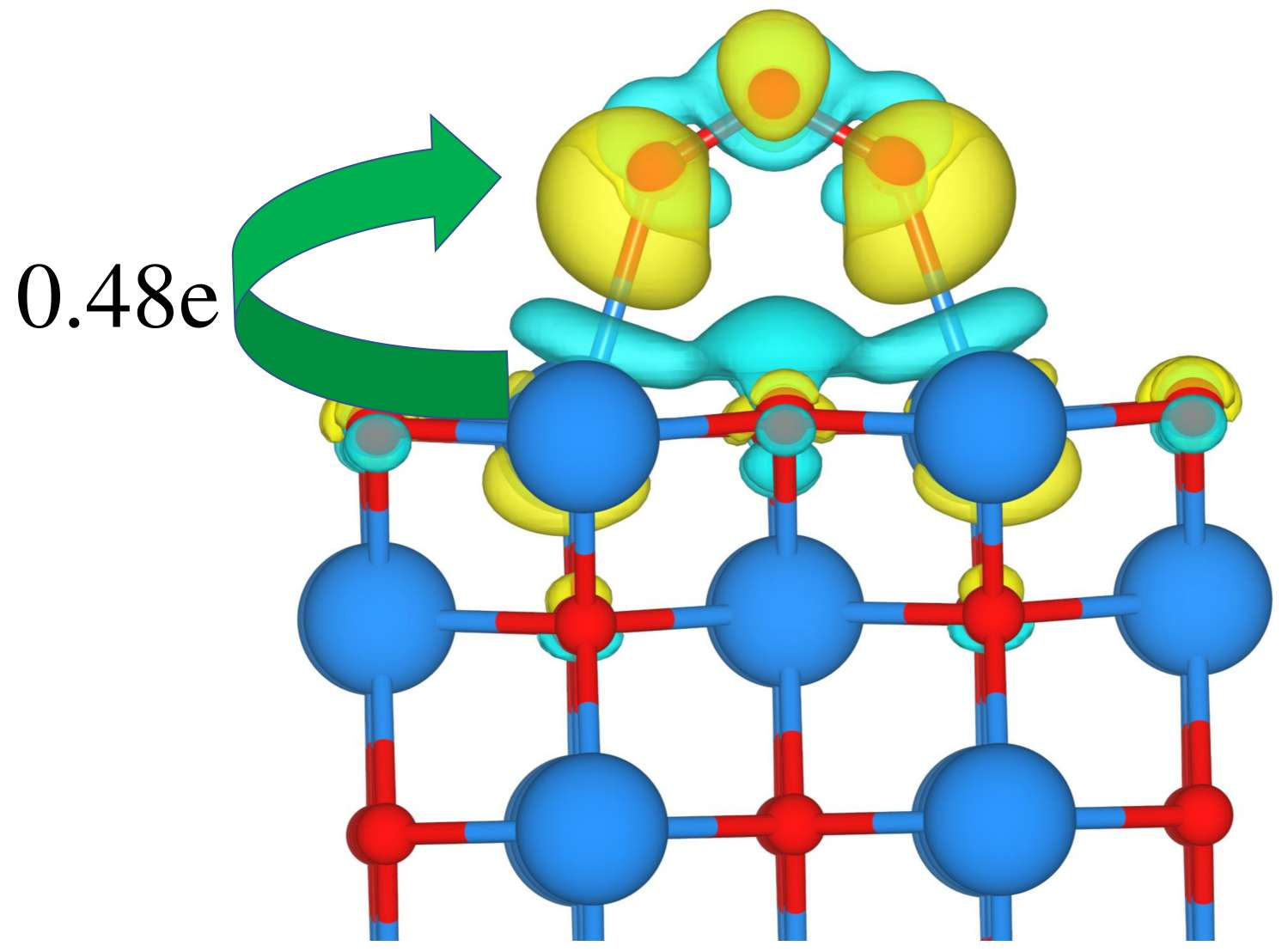}\label{fig: Figure 1e}}
\subfloat[f]{\includegraphics[width= 0.44\textwidth]{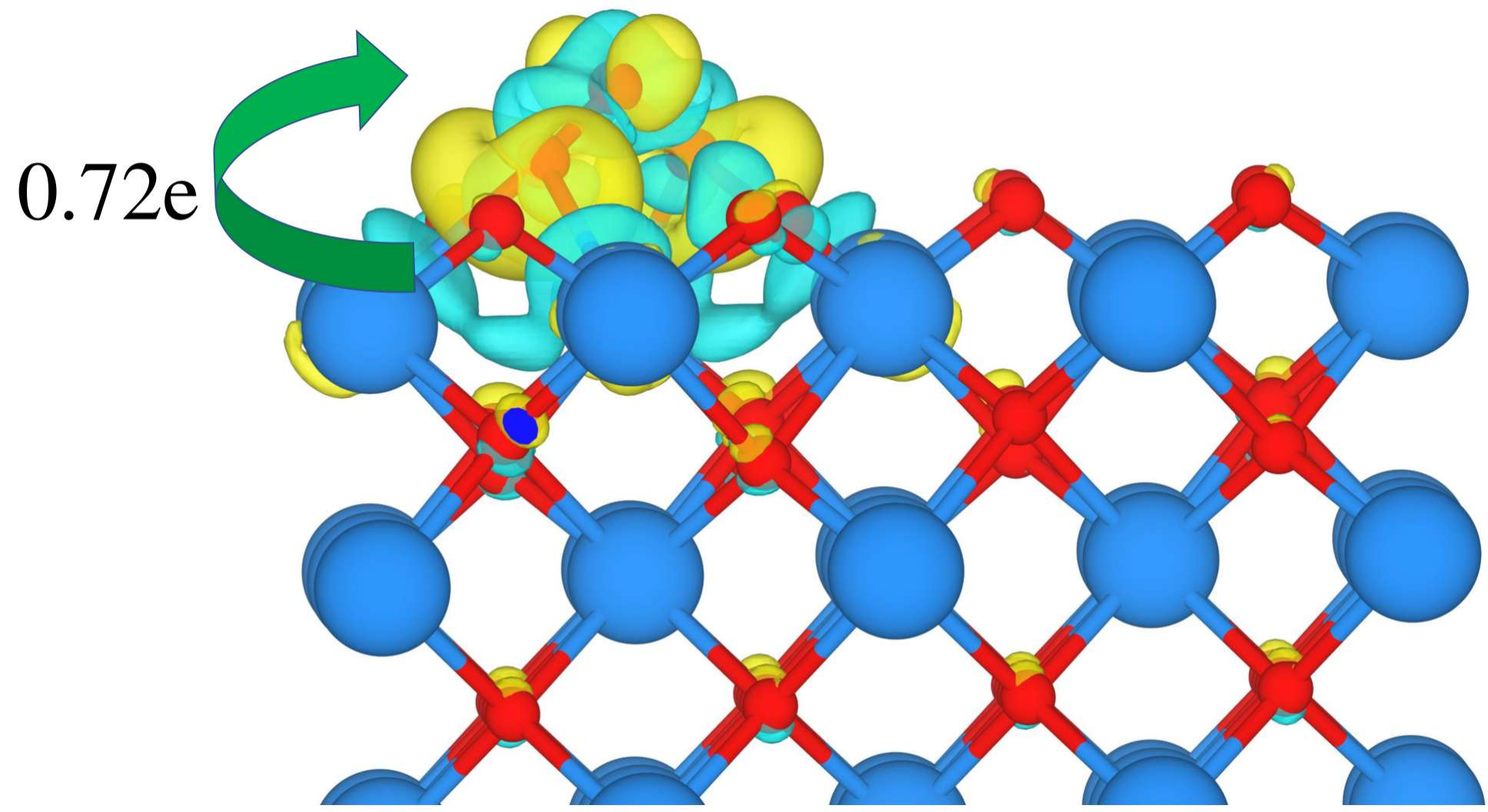}\label{fig: Figure 1f}}
	\caption{Stable binding configurations of O$_3$ on (a) \{111\}, (b) \{110\}, and (c) \{100\} CeO$_2$ surfaces.
Charge density difference plot of O$_3$ interaction with (d) \{111\}, (e) \{110\}, and (f) \{100\} CeO$_2$ surfaces.
The yellow and cyan color isosurfaces show charge accumulation and depletion respectively. Ce and O
atoms are represented by blue and red color spheres respectively. O atoms of O$_3$ are shown by yellow
spheres in order to distinguish it from the surface oxygen atoms. The direction of arrow shows the
direction of electron transfer.}
\label{fig:Figure1}
\end{figure}

O$_3$ gets adsorbed at Ce site of \{111\} surface forming a Ce-O$_{ozone}$ bond of length 2.66 Å with
an adsorption energy of -0.40 eV (Figure \ref{fig:Figure1}\subref{fig: Figure 1a}). As a result of interaction, we notice an 
elongation in the $O_{ozone}-O_{ozone}$ bond length from 1.28 Å in the gaseous phase 
to 1.32 Å in the adsorbed state. The oxygen
atom not bound to the surface atoms has $O_{ozone}-O_{ozone}$ bond length of 1.29 Å. Whereas, on \{110\}
surface O$_3$ binds with an adsorption energy of -0.91 eV and forms a bond with two Ce atoms
of length 2.60 Å resulting in an elongation of O$_{ozone}-O_{ozone}$ bond to 1.33 Å (Figure \ref{fig:Figure1}\subref{fig: Figure 1b}).
On \{100\} surface, O$_3$ binds to the surface through Ce atoms forming a Ce-O$_{ozone}$ bond of 
lengths 2.61-2.63 Å and the
corresponding adsorption energy is -1.04 eV (Figure \ref{fig:Figure1}\subref{fig: Figure 1c}). Interestingly, compared to other surfaces elongation
in O$_{ozone}$-O$_{ozone}$ bond is maximum on \{100\}, the bond length is 1.36 Å after interaction
with the surface. There is a decrease
in the bond angle of O$_3$ upon interaction with the CeO$_2$ \{111\} and \{100\} surfaces and it slightly
increases for \{110\} surface (see Table \ref{table: table1}). The reduction in the bond angle of ozone is more prominent in 
\{100\} and can be attributed to the increased O$_{ozone}$-O$_{ozone}$ bond length which leads to the 
weaker repulsion. Also, the increase in the bond length of ozone will reduce the bond dissociation
energy and facilitate further reactions.

The charge density difference ($\delta\rho$) is calculated using the following relation:
\begin{equation} 
        \delta\rho = \rho_{slab+O_3}-\rho_{slab}-\rho_{O_3} 
\end{equation}
where, $\rho_{slab+O_3}$, $\rho_{slab}$, and $\rho_{O_3}$ are the charge density of slab with O$_3$ adsorbed on it, the CeO$_2$
surface and the isolated O$_3$ molecule in its adsorbed configuration respectively. The charge density
difference analysis shows the accumulation of electrons in Ce-O bond and depletion on the surface
Ce atoms which indicate the electron transfer from the surface to O$_3$ (see Figure \ref{fig:Figure1}\subref{fig: Figure 1d}-\subref{fig: Figure 1f}). Also, we notice 
a charge depletion around one of the oxygen atoms of ozone indicating that it donates electrons
to the other two oxygen atoms of ozone as well. The
Bader charge analysis shows the transfer of electrons from the surface to
the molecule. The electron transfer is maximum on \{100\} surface followed by \{110\} and
\{111\} surface indicating that O$_3$ interacts strongly with \{100\} surface which is also reflected
in the adsorption energy values. In a nutshell, O$_3$ becomes negatively charged with the changed bond
angle and elongated O$_{ozone}$-O$_{ozone}$ bond length upon adsorption on CeO$_2$ surfaces.
The structural parameters and adsorption energies are summarised in Table \ref{table: table1}.

\subsection{O$_3$ dissociation on stoichiometric \{111\}, \{110\}, and \{100\} surfaces}
The adsorption energies of dissociative adsorption of O$_3$ on stoichiometric CeO$_2$ surfaces are 
-0.15 eV, -1.31 eV, and -1.46 eV for the \{111\}, \{110\}, and \{100\} surfaces respectively. The lowest
energy geometries of dissociative O$_3$ on CeO$_2$ surfaces are shown in Figure S1 of supplementary information. O$_3$
dissociates into O$_2$-O and the dissociation occurs during minimization. On \{111\} surface, O$_2$ gets
attached to the surface Ce atom forming two Ce-O$_{ozone}$ bonds of length 2.54 \AA{} and 2.47 Å and other
O$_{ozone}$ atom binds with the surface O and Ce atoms, the corresponding bond lengths are 1.40 Å and
2.55 Å respectively (Figure S1(a) of supplementary information). At \{110\} surface, O$_2$ (of O$_3$) forms a single 
Ce-O$_{ozone}$ bond of length 2.37 Å and O gets attached to the surface Ce and O atoms 
forming a bond of length 2.46 Å and 1.37 Å
respectively (Figure S1(b) of supplementary information). Whereas, on \{100\} surface O$_2$ (of O$_3$) forms Ce-O bond of lengths 2.48-2.68 Å.
O (of O$_3$) gets attached to two Ce atoms and the surface oxygen atom (Figure S1(c) of supplementary information). The structural parameters
and binding energies are given in Table S1 of supplementary information. 

O$_3$ dissociation is thermodynamically more
favorable on \{110\} and \{100\} surfaces as compared to the \{111\} surface. However, the oxygen species formed 
as a result of ozone decomposition should not get stabilised on the surface.
This will make it difficult for the species to leave the surface and the adsorption sites
won't be available to carry out further reaction. As per the adsorption energy, oxygen species formed
as a result of ozone decomposition is quite stable on \{110\} and \{100\} surfaces
as compared to \{111\} which makes \{111\} surface more suitable for ozone decomposition.

\subsection{O$_3$ adsorption on reduced \{111\}, \{110\}, and \{100\} surfaces}
\begin{figure}[ht]
\centering
\subfloat[a]{\includegraphics[width= 0.2\textwidth]{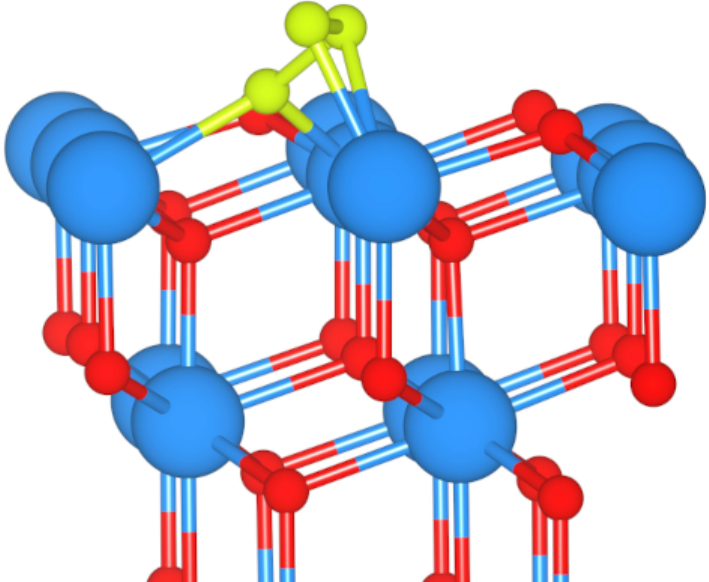}\label{fig: Figure 2a}}
        \hspace{0.3 cm}
\subfloat[b]{\includegraphics[width= 0.29\textwidth]{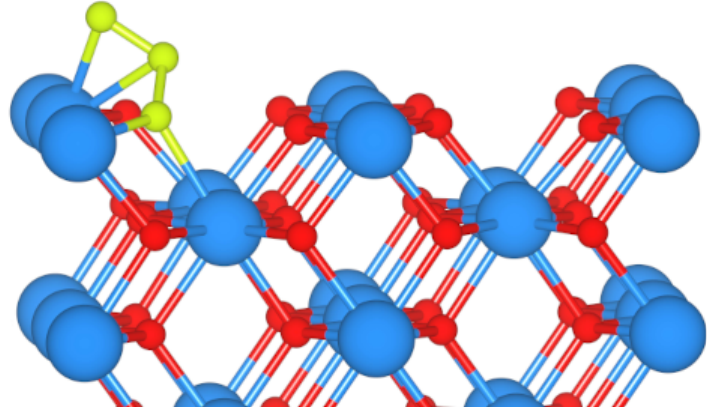}\label{fig: Figure 2b}}
        \hspace{0.3 cm}
        \subfloat[c]{\includegraphics[width= 0.31\textwidth]{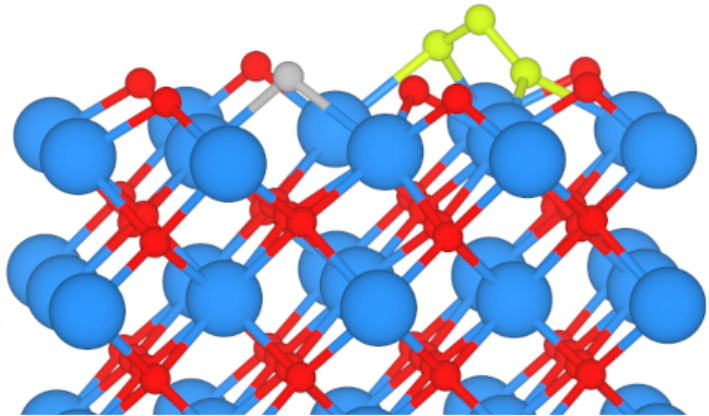}\label{fig: Figure 2c}}

	\caption{Adsorption of O$_3$ on reduced (a) \{111\}, (b) \{110\}, and (c) \{100\} surfaces of CeO$_2$. O$_3$ occupy the oxygen deficient site in \{111\} and \{110\} surfaces. Oxygen deficient site is represented by grey sphere.}
 \label{fig: Figure2}
\end{figure}

As oxygen vacancies act as an active site for ozone adsorption and decomposition,
we have introduced the oxygen vacancies of concentrations 0.25, 0.125, and 0.125 on \{111\},
\{110\}, and \{100\} surfaces respectively. At \{111\} and \{110\} surfaces, O$_3$ occupy the oxygen
deficient site with an adsorption energy of -3.24 eV and -2.65 eV respectively (Figure \ref{fig: Figure2}\subref{fig: Figure 2a} and \subref{fig: Figure 2b}). 
Whereas, on \{100\} surface O$_3$ binds to Ce atoms slightly away from the oxygen-deficient site, 
its adsorption energy is -2.71 eV (Figure \ref{fig: Figure2}\subref{fig: Figure 2c}). The binding energies and structural parameters
are tabulated in Table \ref{table: table2}. We notice a significant elongation
in O$_{ozone}$-O$_{ozone}$ bond length on reduced surfaces as compared to the stoichiometric.
 This indicates that the presence of oxygen vacancies facilitate the decomposition more efficiently 
 as compared to the stoichiometric surfaces.

\begin{figure}[ht]
\centering
\subfloat[a]{\includegraphics[width= 0.2\textwidth]{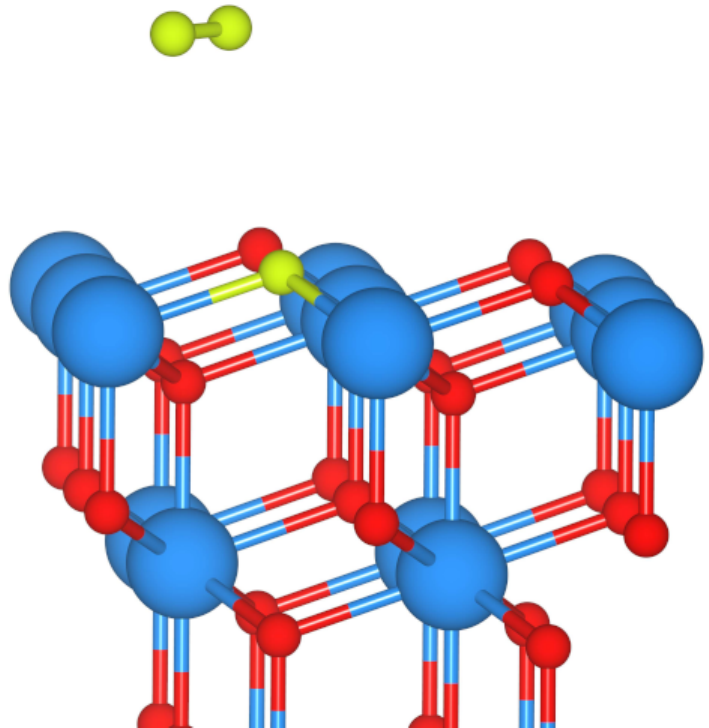}\label{fig: Figure 3a}}
        \hspace{0.3 cm}
\subfloat[b]{\includegraphics[width= 0.29\textwidth]{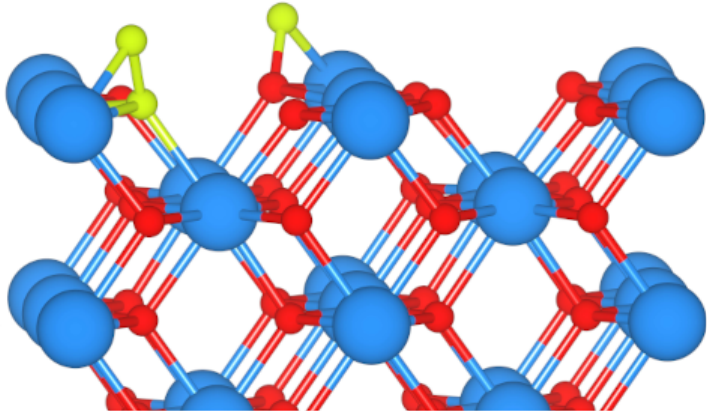}\label{fig: Figure 3b}}
        \hspace{0.3 cm}
        \subfloat[c]{\includegraphics[width= 0.31\textwidth]{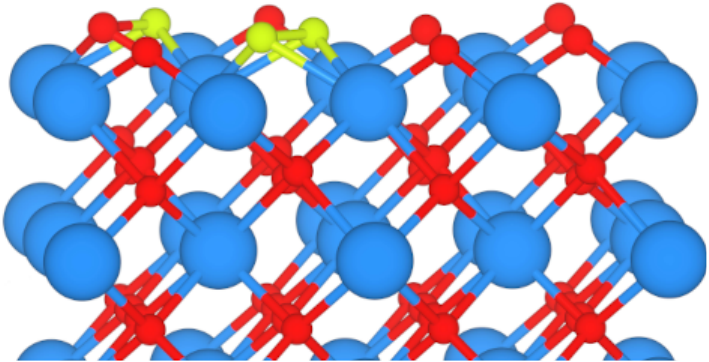}\label{fig: Figure 3c}}
	\caption{Dissociatively adsorbed O$_3$ on reduced (a) \{111\}, (b) \{110\}, and (c) \{100\} surfaces of CeO$_2$.}
 \label{fig: Figure3}
\end{figure}

Considering the dissociative adsorption of O$_3$ on reduced CeO$_2$ surfaces we find that O$_3$
adsorbed in dissociative form is thermodynamically more favorable on \{111\} surface
as compared to \{110\} and \{100\} surfaces (see Table S2 of supplementary information). O$_3$ dissociates into O$_2$-O on \{111\} surface
with O$_{ozone}$ atom occupying the oxygen-deficient site and O$_2$ (of O$_3$) gets stabilized at 3.04 Å
from the surface (Figure \ref{fig: Figure3}\subref{fig: Figure 3a}). It clearly shows the release of oxygen gas after O$_3$ dissociation. Contrary to
\{111\} surface, O$_2$ of O$_3$ occupies the oxygen deficient site and O$_{ozone}$ binds with the surface O
and Ce atoms on both \{110\} and \{100\} surfaces (Figure \ref{fig: Figure3}\subref{fig: Figure 3b} and \subref{fig: Figure 3c}). The adsorption energies of O$_3$ on reduced
CeO$_2$ surfaces reveal that the binding becomes stronger in the presence of oxygen vacancies. 
On \{111\} surface of reduced CeO$_2$, oxygen gas is released spontaneously after dissociation which make 
it a good candidate for ozone decomposition as compared to \{110\} and \{100\} surfaces in the presence
of oxygen vacancies.

\subsection {H$_2$O adsorption on stoichiometric \{111\}, \{110\}, and \{100\} surfaces}
The presence of H$_2$O affect the catalytic activity of O$_3$ decomposition on the oxide surfaces.
We have obtained the most stable adsorption site of H$_2$O both in associative and dissociative state
on \{111\}, \{110\}, and \{100\} surfaces. The adsorption energies and structural geometries are tabulated
in Table S3 of supplementary information. The adsorption energies of H$_2$O adsorbed in the associative form are -0.54 eV, -0.77 eV,
and -0.92 eV on \{111\}, \{110\}, and \{100\} surfaces respectively. H$_2$O interacts strongly with
\{100\} surface and weakly with \{111\} surface, similar trend was also observed for O$_3$ adsorption.
The interaction of H$_2$O with \{111\} surface is slightly stronger in comparison to O$_3$ whereas, on \{110\}
and \{100\} surfaces O$_3$ interaction is stronger than the H$_2$O. H$_2$O prefers to adsorb at Ce site
forming a Ce-O$_{H_2O}$ bond of length 2.61 Å and 2.65 Å on \{111\} and \{110\} surfaces respectively
(Figure S2(a) and (b) of supplementary information).
We notice the formation of one hydrogen bond of length 1.74 Å on \{111\} surface, whereas H$_2$O
forms two hydrogen bonds of length 2.00 Å on \{110\} surface. Contrary to \{111\} and \{110\}
surface, H$_2$O is shared between two Ce atoms forming a Ce-O$_{H_2O}$ bond of length 2.65 Å on \{100\}
surface (Figure S2(c) of supplementary information).

The adsorption energies of H$_2$O adsorbed in dissociative form on \{111\}, \{110\}, and \{100\}
surfaces are -0.52 eV, -1.07 eV, and -1.69 eV respectively. OH binds to the Ce atom and H 
gets attached to the adjacent surface oxygen atom in the most stable configuration on \{111\} and
\{110\} surfaces (Figure S2(d) and (e) of supplementary information). The adsorption energy difference between H$_2$O adsorbed in 
associative and dissociative form on \{111\} surface is 0.02 eV suggesting the co-existence of H$_2$O in both
associative and dissociative state on \{111\} surface. At \{100\} surface, OH occupies the vacant site
created as a result of the removal of half of the surface oxygen atoms and H gets attached to the surface oxygen
atom (Figure S2(f) of supplementary information). The adsorption energies and structural parameters are given in Table S3 of supplementary information. Our adsorption
energies and geometrical parameters are in close agreement with the previous theoretical studies
done by Molinari et al.\cite{Marco.2012}.

\subsection {H$_2$O adsorption on reduced \{111\}, \{110\}, and \{100\} surfaces}
The adsorption energies of H$_2$O adsorbed in associative form on reduced CeO$_2$ \{111\}, \{110\}, and
\{100\} surfaces are -1.21 eV, -1.20 eV, and -1.00 eV respectively. H$_2$O forming bond with the
surface Ce atom is the most stable binding configuration and there is a hydrogen bond
formation on all the considered surfaces (see Figure S3(a)-(c) of supplementary information). 
In case of H$_2$O adsorbed in dissociative form, OH prefers
to occupy the oxygen-deficient site and O$_{water}$ gets attached to the surface oxygen atom. On \{110\}
surface, OH forms a hydrogen bond of length 1.81 Å, there is no hydrogen bond formation on any
other surfaces as shown in Figure S3(d)-(f) of supplementary information. The adsorption energies of both O$_3$ and H$_2$O 
are more negative on reduced CeO$_2$ surfaces indicating the strong interaction between vacancies 
and adsorbates. The results are in agreement with the previous work done by
Molinari et al. \cite{Marco.2012}, Fronzi et al. \cite{Fronzi.2009}, and 
Watkins et al. \cite{Watkins.2007}.  
However, the presence of oxygen vacancies favors the binding of O$_3$ as
compared to that of H$_2$O. 
The adsorption energies and geometrical parameter are summarised in
Table S4 of supplementary information.

\subsection{Competition between H$_2$O and O$_3$ on CeO$_2$ \{111\} surface: Coadsorption study} 
\begin{figure}[ht]
\centering
\subfloat[a]{\includegraphics[width= 0.3\textwidth]{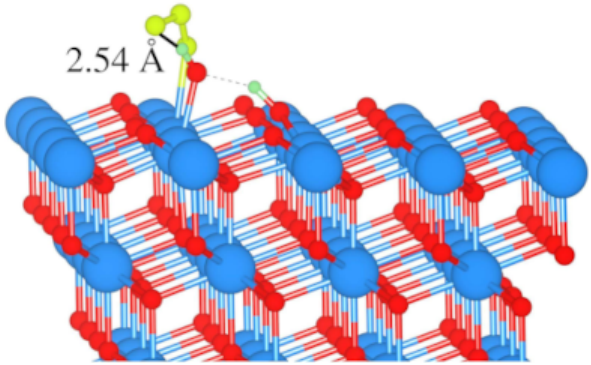}\label{fig: Figure 4a}}
        \hspace{0.3 cm}
\subfloat[b]{\includegraphics[width= 0.3\textwidth]{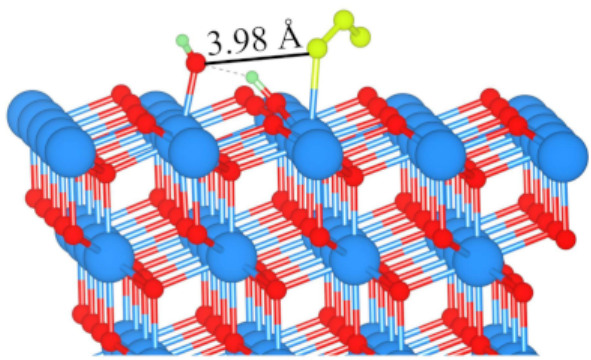}\label{fig: Figure 4b}}
        \hspace{0.3 cm}
        \subfloat[c]{\includegraphics[width= 0.3\textwidth]{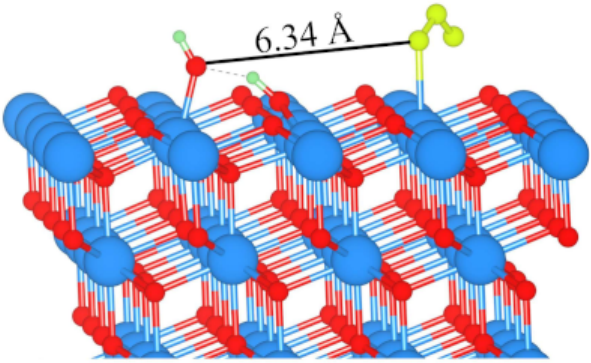}\label{fig: Figure 4c}}\\
	\subfloat[d]{\includegraphics[width= 0.3\textwidth]{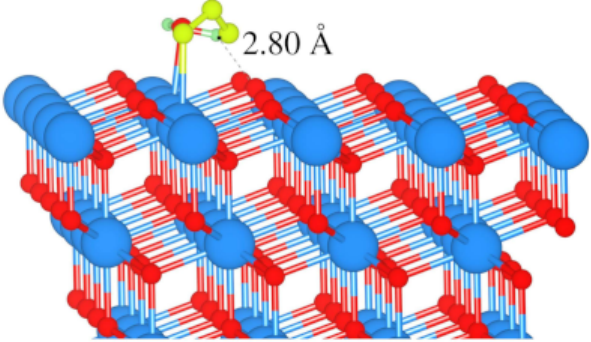}\label{fig: Figure 4d}}
        \hspace{0.3 cm}
\subfloat[e]{\includegraphics[width= 0.3\textwidth]{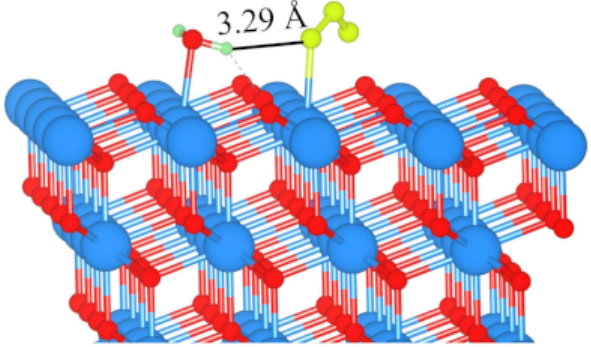}\label{fig: Figure 4e}}
        \hspace{0.3 cm}
        \subfloat[f]{\includegraphics[width= 0.3\textwidth]{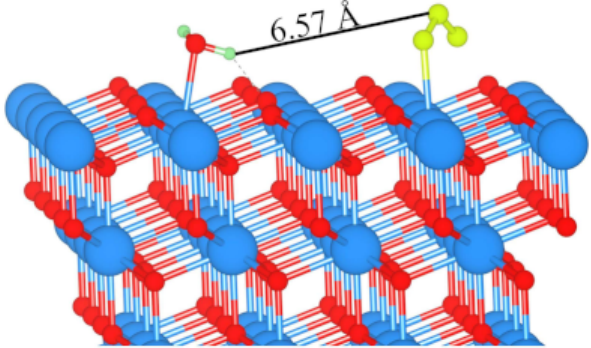}\label{fig: Figure 4f}}\\
	 \caption{Structures of H-OH and O$_3$ (a)-(c) and H$_2$O and O$_3$ (d)-(f) on stoichiometric CeO$_2$\{111\} surface, the black solid lines show the distance between H-OH/H$_2$O and O$_3$. Dotted lines are used to show hydrogen bond formation.}
    \label{fig: Figure4}
\end{figure}

From our previous results, it is clear that both O$_3$ and H$_2$O are strongly attracted to the surface
and prefers same binding sites for adsorption. There will be competitive adsorption between 
O$_3$ and H$_2$O on the CeO$_2$ surfaces.
As CeO$_2$\{111\} surface is the most stable surface of all, we have investigated the coadsorption of
O$_3$ and H$_2$O on stoichiometric CeO$_2$\{111\} surface. We have looked into the effect 
of H$_2$O and H-OH on the adsorption of ozone as both H$_2$O and H-OH
are equally probable on \{111\} surface. 
We have considered $2\times2$ conventional unit cells for the coadsorption of H$_2$O and O$_3$ on
stoichiometric \{111\} surface. For a better comparison of O$_3$ adsorption energy in the presence
and absence of H$_2$O, we have obtained the adsorption energy of O$_3$ on $2\times2$ conventional unit
cells and got the adsorption energy of O$_3$ to be -0.43 eV. The adsorption energy is changed by
0.03 eV when the coverage of O$_3$ is reduced indicating that the O$_3$ coverage has a little effect
on its binding on stoichiometric \{111\} surface of CeO$_2$. For the same coverage, the adsorption energy
of H-OH is -0.64 eV, it is changed by 0.07 eV indicating that H-OH adsorption energy
is coverage dependent as reported by Molinari \textit{et al.} \cite{Marco.2012}. 
To look into the effect of H-OH on the
adsorption energy of O$_3$ we varied the distance between O$_3$ and H-OH (see Figure \ref{fig: Figure4}\subref{fig: Figure 4a}-\subref{fig: Figure 4c}) and 
calculated the adsorption energy of O$_3$ using the following relation:
\begin{equation}
	E_{ads.} = E_{slab+H-OH/H_2O+O_3}-(E_{slab+H-OH/H_2O}+E_{O_3})	
\end{equation}

Where, E$_{slab+H-OH/H_2O+O_3}$, is the energy of H-OH/$H_2O$ and O$_3$ adsorbed on slab, E$_{slab+H-OH/H_2O}$ is the
energy of H-OH and H$_2$O adsorbed on slab and E$_{O_3}$ is the energy of isolated O$_3$ in
the gaseous phase. The adsorption energy of O$_3$ in the presence of H-OH is -0.26 eV when kept
at a distance of 2.54 \AA{} and its 
-0.43 eV in the absence of H-OH. Also, O$_3$ moves further away from the surface and binds at a
distance of 2.75 Å (Ce-O$_{ozone}$ bond length) from the surface in the presence of H-OH indicating 
weaker interaction between O$_3$ and surface (Table \ref{table: table3}). This suggests that the 
presence of H-OH weakens the binding of
O$_3$ on stoichiometric \{111\} surface. However, when O$_3$ is moved away from H-OH i.e. at a
distance of 6.34 Å, there is a negligible interaction between H-OH and O$_3$, its adsorption energy
is -0.38 eV which is close to its adsorption energy value of -0.43 eV in the absence of H-OH. At
an intermediate distance of 3.98 \AA{}, the binding energy of ozone is -0.27 eV indicating
there is still an interaction between O$_3$ and H-OH.
Charges and other structural parameters of O$_3$ are similar to that in the absence of H-OH at 6.34 \AA{} (see Table \ref{table: table3}). O$_3$ becomes negatively charged and H-OH becomes slightly positively charged as a result of
coadsorption. 

\begin{figure}[ht]
\centering
\subfloat[a]{\includegraphics[width= 0.46\textwidth]{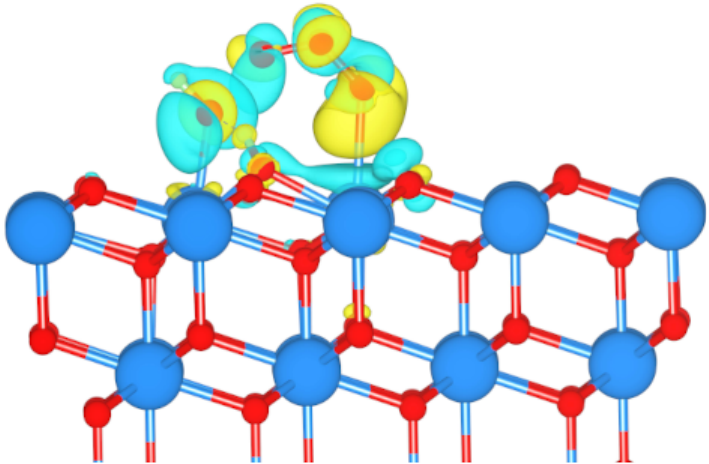}\label{fig: Figure 5a}}
        \hspace{0.5 cm}
\subfloat[b]{\includegraphics[width= 0.48\textwidth]{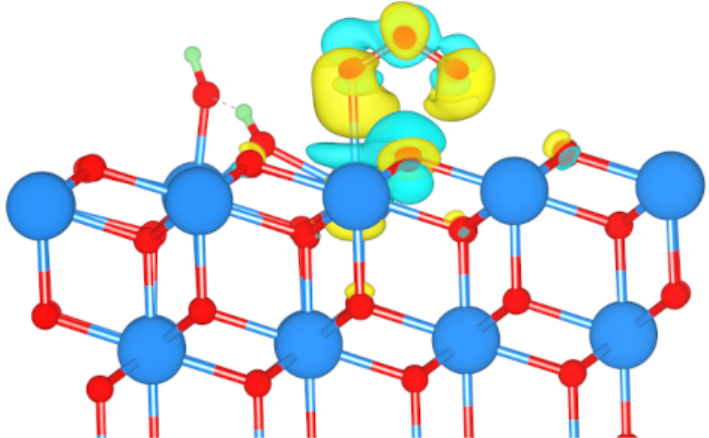}\label{fig: Figure 5b}}\\
\subfloat[c]{\includegraphics[width= 0.46\textwidth]{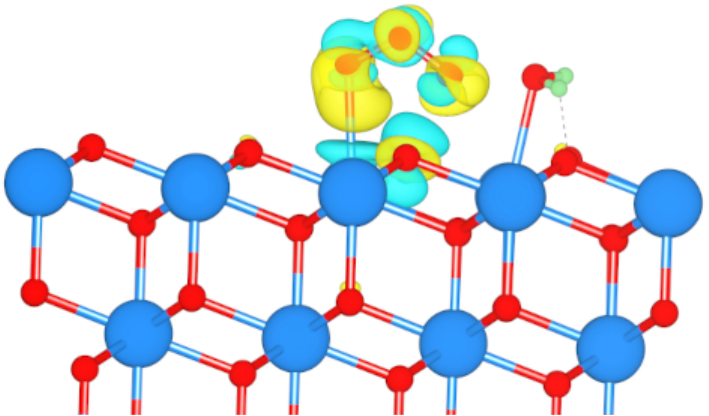}\label{fig: Figure 5c}}
        \hspace{0.5 cm}
\subfloat[d]{\includegraphics[width= 0.48\textwidth]{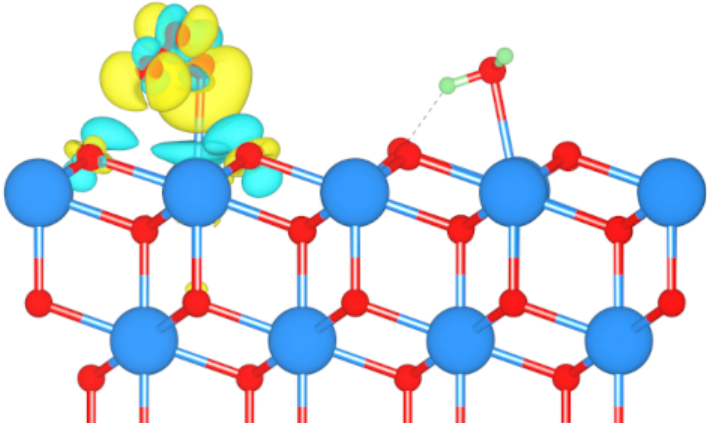}\label{fig: Figure 5d}}
         \caption{Charge density difference plots of O$_3$ interaction with \{111\} surface of CeO$_2$ in the presence of H-OH when kept at a distance of (a) 2.54 Å, and (b) 6.34 Å from O$_3$ and when the distance between
	 H$_2$O and O$_3$ is (c) 2.80 \AA{} and (d) 6.57 \AA{}. The yellow and cyan color
isosurfaces show charge accumulation and depletion respectively.}
 \label{fig: Figure5}
\end{figure}

We further calculated the charge density difference ($\delta\rho)$ plot of two configurations in which O$_3$ is
at distance of 2.54 Å and 6.34 Å from H-OH using the following relation:
\begin{equation} 
	\delta\rho = \rho_{slab+H-OH/H_{2}O+O_3}-\rho_{slab+H-OH/H_{2}O}-\rho_{O_3}, 
\end{equation}
where $\rho_{slab+H-OH/H_{2}O+O_3}$, $\rho_{slab+H-OH/H_{2}O}$, and $\rho_{O_{3}}$ are
the charge density of slab with H-OH/H$_2$O and O$_3$ adsorbed on it, the slab with H-OH/H$_2$O 
adsorbed on it, and the isolated O$_3$ molecule in its adsorbed configuration.
We notice an electron accumulation along Ce-O$_{ozone}$ bond and electron depletion on surface atoms
suggesting the electron transfer from the surface to O$_3$. Also, there is a charge re-distribution on the
H-OH when it is close to O$_3$ (at a distance of 2.54 Å), whereas it remains unaffected
when it is moved further away from O$_3$ as shown in Figure \ref{fig: Figure5}\subref{fig: Figure 5a} and \subref{fig: Figure 5b}. Our analysis confirms that 
there is an interaction
between H-OH and O$_3$ when placed close to each other which clearly affects the binding and
geometry of O$_3$ (see Table \ref{table: table3}).

We also performed the analysis of H$_2$O and O$_3$ co-adsorption on \{111\} surface of CeO$_2$.
The adsorption energy of H$_2$O for the same coverage is -0.57 eV.
In the presence of H$_2$O, the binding energies of ozone obtained using 
equation 3 are -0.36 eV, -0.36 eV, and -0.39 eV 
when kept at a distance of 2.80 \AA{}, 3.29 \AA{}, and 6.57 \AA{} respectively from H$_2$O
 (Figure \ref{fig: Figure4}\subref{fig: Figure 4d}-\subref{fig: Figure 4f}).
There is a smaller change in the binding energies of ozone with its distance from H$_2$O suggesting 
that there is a negligible interaction between H$_2$O and O$_3$. We performed the Bader charge 
analysis for the two cases: one in which O$_3$ is at a distance of 2.80 \AA{} and other at
6.57 \AA{} from H$_2$O. There is an electron transfer from 
surface to O$_3$ molecule as per the Bader charge analysis (Table \ref{table: table4}). Also, the charge density difference
obtained using equation 4 reveals the electron accumulation along Ce-O$_{ozone}$ bond and depletion along the 
surface Ce atom. There is no charge re-distribution on the H$_2$O irrespective of its distance
from H$_2$O molecule which further confirms the
negligible interaction between H$_2$O and O$_3$ (Figure \ref{fig: Figure5}\subref{fig: Figure 5c} and \subref{fig: Figure 5d}).

The effect of H-OH on O$_3$ adsorption energy is much more as compared to H$_2$O. This can be attributed
to the strong interaction of H-OH with the surface as compared to the surface-H$_2$O interaction.
As a result of which competitive adsorption between H-OH and O$_3$ weakens its interaction with the surface and
hence the adsorption energy of O$_3$ becomes more positive in the presence of H-OH.

\begin{figure}[ht]
\centering
\subfloat[a]{\includegraphics[width= 0.47\textwidth]{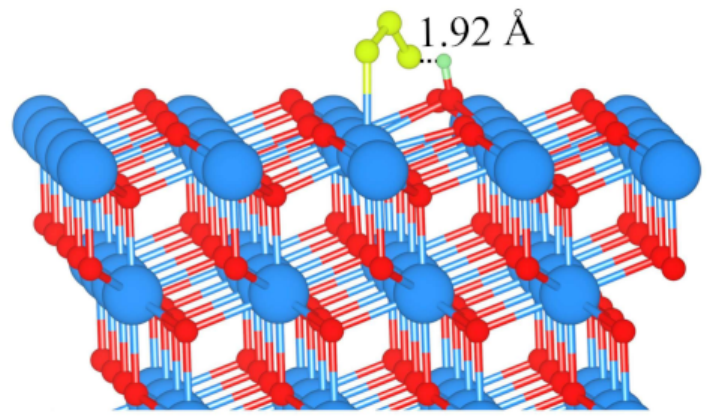}\label{fig: Figure 6a}}
        \hspace{0.45 cm}
\subfloat[b]{\includegraphics[width= 0.47\textwidth]{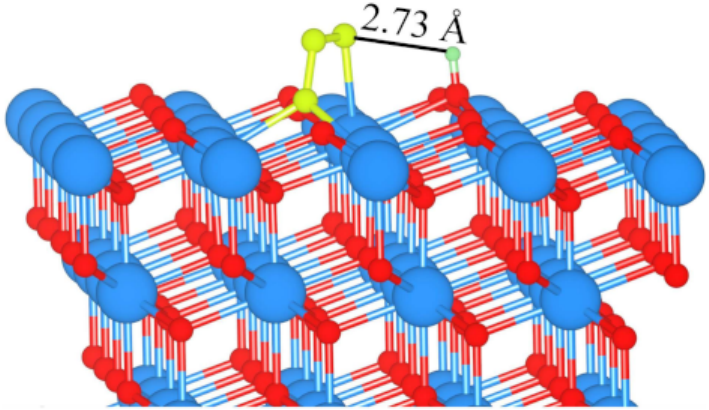}\label{fig: Figure 6b}}
	 \caption{Stable binding configuration of OH and O$_3$ on the reduced \{111\} CeO$_2$ surface for the vacancy concentration (a) 0.0625, and (b) 0.125. The dotted line shows the hydrogen bond.}
  \label{fig: Figure6}
\end{figure}

As per the adsorption energy, the H-OH is more favorable on reduced \{111\} surface as
compared to H$_2$O and so we consider the coadsorption of O$_3$ and OH on reduced CeO$_2$ surface.
In this case we are considering the effect of isolated hydroxyl group on ozone adsorption.
For simulation of reduced CeO$_2$\{111\} surface we have considered $2\times2$ conventional unit cells and
removed one oxygen atom introducing an oxygen vacancy of concentration 0.0625. Also, the
adsorption energy of O$_3$ on \{111\} surface with vacancy concentration of 0.0625 in the absence
of H$_2$O is -2.85 eV. O$_3$ prefers to occupy the oxygen deficient site in the absence of H$_2$O. In
case of co-adsorption, there is one oxygen deficient site which gets occupied by OH. Since there is
no available oxygen deficient site, O$_3$ binds to the surface Ce atom forming a Ce-O bond of
length 2.45 Å (Figure \ref{fig: Figure6}\subref{fig: Figure 6a}). There is a hydrogen bond formation between OH and O$_3$ of length 1.92 Å. The
adsorption energy of O$_3$ in the presence of OH is calculated using the following relation:
\begin{equation}
	E_{ads.}(O_3) = E_{slab+OH+O_3}-(E_{slab+OH}+E_{O_3})
\end{equation}

Where, E$_{slab+OH+O_3}$, is the energy of OH and O$_3$ adsorbed on a slab, E$_{slab+OH}$ is the energy of OH
adsorbed on slab, and E$_{O_3}$ is the energy of O$_3$ in the isolated gaseous phase. The adsorption energy
of O$_3$ in the presence of OH is -1.58 eV i.e., its binding becomes less negative compared to its
adsorption energy of -2.85 eV in the absence of H$_2$O. The stable binding geometry and
geometrical parameters obtained because of coadsorption of O$_3$ and OH is shown in Figure \ref{fig: Figure6}\subref{fig: Figure 6a} and tabulated
in Table \ref{table: table5}. We increased the concentration of oxygen vacancies to 0.125 on the surface by removing two
surface oxygen atoms. In this case the adsorption energy of
ozone in the absence of OH is -2.98 eV. OH occupies the oxygen deficient site and O$_3$ also has
one deficient site for adsorption. O$_3$ fills the vacancy site created as a result
oxygen vacancy formation (see Figure \ref{fig: Figure6}\subref{fig: Figure 6b}), its binding energy is -2.69 eV 
obtained using equation 5. In this case also, we do notice that the presence of OH weakens the interaction between O$_3$ and surface.  

\begin{figure}[ht]
	\centering
\includegraphics[width=0.7\textwidth]{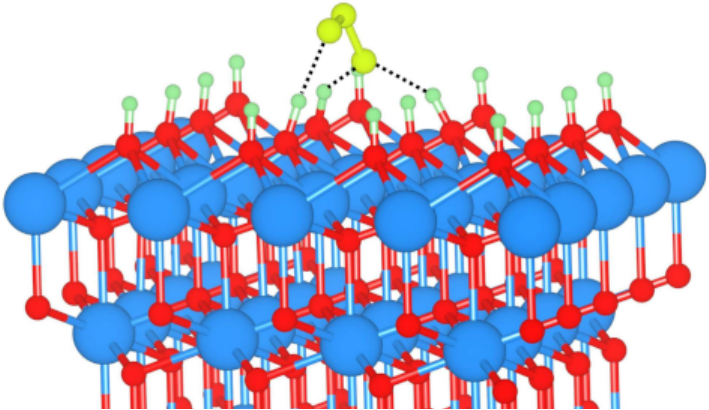}
	\caption{Stable binding configuration of O$_3$ on hydroxylated \{111\} surface. The dotted lines show
the hydrogen bonds.}
\label{fig: Figure7}
\end{figure}

On the basis of above result we looked into the effects of hydroxylated surface on ozone adsorption. 
The adsorption energy of O$_3$ is calculated using the following relation: 
\begin{equation}
	E_{ads.}(O_3) = E_{slab+H+O_3}-(E_{slab+H}+E_{O_3})
\end{equation}
Where, E$_{slab+H+O_3}$ is the energy of O$_3$ adsorbed on hydroxylated slab, E$_{slab+H}$ is the energy of
hydroxylated slab, and E$_{O_3}$ is the energy of O$_3$ in the isolated gaseous phase. On the hydroxylated
surface, O$_3$ stays at the top of surface forming three hydrogen bonds of lengths 1.92, 1.86, and
1.78 Å (shown in Figure \ref{fig: Figure7}). Our results suggest that O$_3$ can not come closer to surface and
forms chemical bond but the presence of hydrogen bonds are strengthening the interaction between
surface and O$_3$ leading to its adsorption energy of -1.74 eV.  We find that the presence of OH 
interfers with the adsorption of ozone on reduced \{111\} surfaces by blocking the 
oxygen-deficient site which is considered to be the active site for ozone adsorption
and decomposition.

\section{Conclusions}
This study explores the adsorption of H$_2$O and O$_3$ on various surfaces of CeO$_2$ and investigates their stability and binding properties. The results show that the stoichiometric \{110\} and \{100\} surfaces exhibits stronger binding with O$_3$ and weaker binding with H$_2$O, whereas the stable \{111\} surface demonstrates slightly stronger binding with H$_2$O compared to O$_3$. On the other hand, the \{111\}, \{110\} and \{100\} surfaces show strong binding with O$_3$, particularly in the presence of oxygen vacancies. The preferred adsorption site on the stoichiometric CeO$_2$ surface is the Ce site, but on the reduced surface, the adsorbate either binds near an oxygen-deficient site or occupies it.

The presence of H-OH weakens the binding of O$_3$, indicating water poisoning, and modifies its structure. O$_3$ forms a longer Ce-O bond when bound to the surface Ce atom, indicating a weaker interaction and reduced electron transfer. The adsorption energy and geometrical parameters of O$_3$ remain unchanged when placed farther away from H-OH. There is minimal interaction between H$_2$O and O$_3$, meaning the presence of H$_2$O does not affect ozone adsorption.

The presence of OH reduces the negative adsorption energy of O$_3$, especially when oxygen vacancies are increased, leading to weaker ozone interaction. A fully hydroxylated surface prevents O$_3$ from forming chemical bonds with surface atoms, but hydrogen bonding between O$_3$ and hydrogen atoms provides stability.

In conclusion, this research highlights the importance of moisture control to prevent water poisoning and optimize the performance of CeO$_2$-based catalysts in industrial applications aimed at reducing ground-level ozone pollution. Heat treatment can be employed to reduce vacancy concentration, increase available adsorption sites for O$_3$, and minimize competitive adsorption by H$_2$O.

\section*{Acknowledgments}
This work was supported financially by the Fundamental Research Program (PNK9400) of the Korea Institute of Materials Science (KIMS). The authors also acknowledge the support by the Korea Institute of Science and Technology, GKP (Global Knowledge Platform, Grant number 2V6760) project of the Ministry of Science, ICT and Future Planning.

\bibliographystyle{prsty}
\bibliography{citation}

\section{Tables and Graphics}
\begin{table}[ht]
\centering
	\caption{Structural parameters, adsorption energies (E$_{ads.})$, and Bader charges of O$_3$
	 adsorbed on $CeO_2$ surfaces. Oxygen atom of $O_3$ is represented by $O_{ozone}$.}
        \label{table: table1}
    \begin{tabular}{ccccc}
\noalign{\smallskip} \hline\hline   \noalign{\smallskip}
	    Parameters & Free O$_3$ & CeO$_2$\{111\}  & CeO$_2$\{110\} & CeO$_2$\{100\}    \\
\hline 
	    $E_{ads.}$ (eV)& - & -0.40 & -0.91 & -1.04 \\
	    Q(O$_3$)($|e|$)  & 0 & -0.25 & -0.48 & -0.72 \\
	    O$_{ozone}$-O$_{ozone}$ (in \AA{}) & 1.28, 1.28 & 1.32, 1.29 & 1.33, 1.33 & 1.36, 1.36 \\
	    O$_3$ bond angle ($^\circ$)  & 118.2 & 116.4 & 118.4 & 113.9 \\
	    Ce-O$_{ozone}$ (\AA{}) & - & 2.66 & 2.60 & 2.61-2.63 \\
\noalign{\smallskip} \hline\hline  \noalign{\smallskip}
\end{tabular}
\end{table} 

\begin{table}[ht]
\centering
	\caption{Structural parameters, and adsorption energies (E$_{ads.}$) of O$_3$
          on reduced $CeO_2$ surfaces. Oxygen atom of $O_3$ is represented by $O_{ozone}$.}
           \label{table: table2}
    \begin{tabular}{ccccc}
\noalign{\smallskip} \hline\hline   \noalign{\smallskip}
            Parameters & Free O$_3$ & CeO$_2$\{111\}  & CeO$_2$\{110\} & CeO$_2$\{100\}    \\
\hline
            $E_{ads.}$ (eV)& - & -3.24 & -2.65 & -2.71 \\
            O$_{ozone}$-O$_{ozone}$ (\AA{}) & 1.28, 1.28 & 1.54, 1.39 & 1.56, 1.39 & 1.44, 1.46 \\
            O$_3$ bond angle ($^\circ$)  & 118.2 & 109.0 & 110  & 110.2 \\
	    Ce-O$_{ozone}$ (\AA{}) & - & 2.29-2.51 & 2.33-2.77  & 2.40-2.54 \\
\noalign{\smallskip} \hline\hline  \noalign{\smallskip}
\end{tabular}
\end{table}

\begin{table}[ht]
\centering
	\caption{Adsorption energies (E$_{ads.}$) and structural parameters of H-OH and O$_3$ coadsorbed on \{111\} surface
	of CeO$_2$, when both are at a distance (d) of 2.54 \AA{} and 6.34 \AA{} from each other.}
  \label{table: table3}
    \begin{tabular}{ccc}
\noalign{\smallskip} \hline\hline   \noalign{\smallskip}
	    Parameters &  H-OH    \\
	       & d = 2.54 \AA{} & d = 6.34 \AA{} \\
               \hline
            $E_{ads.}$ (eV)& -0.26 & -0.38   \\
	    Q(O$_3$)($|e|$)  & -0.20 & -0.25 \\
	    Q(H-OH)($|e|$) & 0.097 & 0.01 \\
	    O$_{ozone}$-O$_{ozone}$ (\AA{}) & 1.33, 1.28 & 1.32, 1.29 \\
	    O$_3$ bond angle ($^\circ$) & 118.16 & 116.5 \\
	    Ce-O$_{ozone}$ (\AA{}) & 2.75 & 2.66 \\
\noalign{\smallskip} \hline\hline  \noalign{\smallskip}
\end{tabular}
\end{table}

\begin{table}
\centering
        \caption{Adsorption energies (E$_{ads.}$) and structural parameters of H$_2$O and O$_3$ coadsorbed on \{111\} surface
        of CeO$_2$, when both are at a distance (d) of 2.80 \AA{} and 6.57 \AA{} from each other.}
         \label{table: table4}
    \begin{tabular}{ccc}
\noalign{\smallskip} \hline\hline   \noalign{\smallskip}
            Parameters & H$_2$O     \\
     & d = 2.80 \AA{} & d = 6.57 \AA{} \\
               \hline
            $E_{ads.}$ (eV) & -0.36  & -0.39    \\
            Q(O$_3$)($|e|$)  & -0.25 & -0.26  \\    
            Q(H$_2$O)($|e|$) & 0.01  & 0.001 \\
            O$_{ozone}$-O$_{ozone}$ (\AA{}) & 1.32, 1.29  & 1.32, 1.29  \\
            O$_3$ bond angle ($^\circ$) & 115.4 & 116.40  \\
            Ce-O$_{ozone}$ (\AA{}) & 2.70  & 2.66  \\
\noalign{\smallskip} \hline\hline  \noalign{\smallskip}
\end{tabular}
\end{table}

\begin{table}
\centering
	\caption{Adsorption energies (E$_{ads.}$) and structural parameters of OH and O$_3$ coadsorbed on \{111\} surface
        of CeO$_2$ at a vacancy concentrations x=0.0625 and 0.125.}
         \label{table: table5}
    \begin{tabular}{ccc}
\noalign{\smallskip} \hline\hline   \noalign{\smallskip}
	    Parameters & x=0.0625 & x=0.125     \\
               \hline
            $E_{ads.}$ (eV) & -0.58   & -2.69     \\
	    O$_{ozone}$-O$_{ozone}$ (\AA{}) & 1.36, 1.36 & 1.47, 1.32       \\
            O$_3$ bond angle ($^\circ$) & 114.24 & 111.4  \\
            Ce-O$_{ozone}$ (\AA{}) & 2.45  & 2.58-2.83  \\
\noalign{\smallskip} \hline\hline  \noalign{\smallskip}
\end{tabular}
\end{table}

\end{document}


\begin{center}
\begin{table}[h!]
\centering
        \caption*{Table S1: Structural parameters, and adsorption energies (E$_{ads.}$) of O$_3$
         adsorbed in dissociative form (O$_2$-O) on CeO$_2$ surfaces.}
    \begin{tabular}{cccc}
\noalign{\smallskip} \hline\hline   \noalign{\smallskip}
            Parameters &  CeO$_2$\{111\}  & CeO$_2$\{110\} & CeO$_2$\{100\}    \\
\hline
            E$_{ads.} (eV)& -0.15 & -1.31 & -1.46  \\
            Ce-O$_2$ (of ozone)(\AA{})  & 2.54, 2.57 & 2.37 & 2.48-2.68 \\
            Ce-O$_{ozone}$ (\AA{}) & 2.54 & 2.46 & 2.40 \\
            O-O$_{ozone}$ (\AA{}) & 1.40 & 1.37 & 1.37 \\
\noalign{\smallskip} \hline\hline  \noalign{\smallskip}
\end{tabular}
\end{table}

\begin{table}[h!]
\centering
   \caption*{Table S2: Adsorption energies (E$_{ads.}$) of O$_3$ adsorbed on reduced CeO$_2$ surfaces in a dissociative form.}
    \begin{tabular}{ccc}
\noalign{\smallskip} \hline\hline   \noalign{\smallskip}
            Surfaces & & Adsorption energies (eV)    \\
\hline
           \{111\} & O$_2$-O & -3.61 \\
            \{110\} & O$_2$-O & -2.53 \\
               \{100\} & O$_2$-O & -3.01 \\

\noalign{\smallskip} \hline\hline  \noalign{\smallskip}
\end{tabular}
\end{table}

\begin{table}[h!]
\centering
        \caption*{Table S3: Adsorpton energies (E$_{ads.}$) and structural parameters of H$_2$O adsorbed in associative and dissociative form on CeO$_2$ surfaces ($^\ast$ denotes hydrogen bond).}
    \begin{tabular}{cccc}
\noalign{\smallskip} \hline\hline   \noalign{\smallskip}
            Parameters &  CeO$_2$\{111\}  & CeO$_2$\{110\} & CeO$_2$\{100\}    \\
\hline
               & & H$_2$O & \\
               \hline
            E$_{ads.} (eV)& -0.54 & -0.77 & -0.92  \\
            Ce-O$_{water}$ (\AA{})  & 2.61 & 2.65 & 2.65 \\
            H-O$^\ast$ (\AA{}) & 1.74 & 2.00, 2.10 & - \\
            \hline
            & & H-OH & \\
            \hline
            E$_{ads.}$ (eV) & -0.52 & -1.07 & -1.69 \\
            Ce-OH (\AA{}) & 2.23 & 2.13 & 2.32, 2.36 \\
            H-O$^\ast$ & 1.62 & 1.91 & - \\
\noalign{\smallskip} \hline\hline  \noalign{\smallskip}
\end{tabular}
\end{table}

\begin{table}[h!]
\centering
        \caption*{Table S4: Adsorpton energies (E$_{ads.}$) and structural parameters of H$_2$O adsorbed in associative and dissociative form on reduced CeO$_2$ surfaces ($^\ast$ denotes hydrogen bond).}
    \begin{tabular}{cccc}
\noalign{\smallskip} \hline\hline   \noalign{\smallskip}
            Parameters &  CeO$_2$\{111\}  & CeO$_2$\{110\} & CeO$_2$\{100\}    \\
\hline
               & & H$_2$O & \\
               \hline
            E$_{ads.} (eV)& -1.21 & -1.20 & -1.00  \\
            Ce-O$_{water}$ (\AA{})  & 2.66 & 2.69 & 2.71 \\
            H-O$^\ast$ (\AA{}) & 1.61 & 1.61 & 1.67, 1.71 \\
            \hline
            & & H-OH & \\
            \hline
            E$_{ads.}$ (eV) & -2.81 & -2.16 & -2.40 \\
            Ce-OH (\AA{}) & 2.59, 2.59 & 2.61, 2.65 & 2.40, 2.57 \\
            H-O$^\ast$ & - & 1.81 & - \\
\noalign{\smallskip} \hline\hline  \noalign{\smallskip}
\end{tabular}
\end{table}
\clearpage
\begin{figure}[h!]
\centering
\subfloat[a]{\includegraphics[width= 0.2\textwidth]{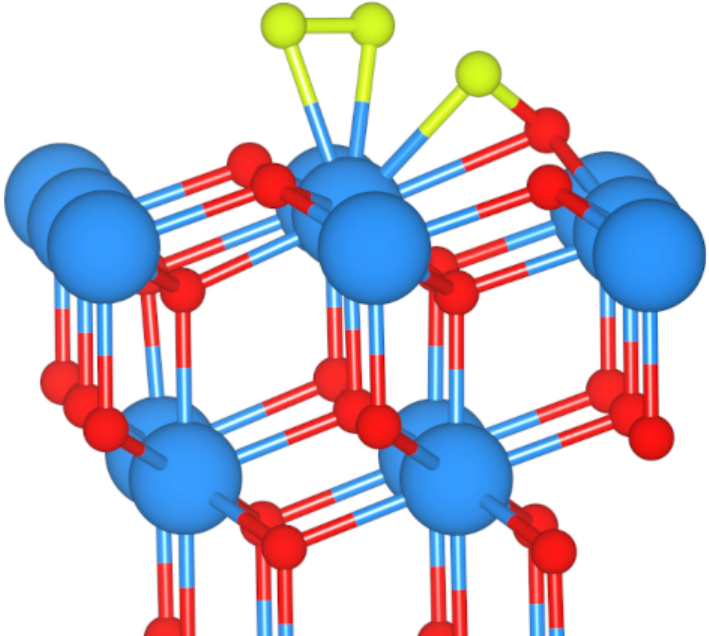}}
        \hspace{0.3 cm}
\subfloat[b]{\includegraphics[width= 0.29\textwidth]{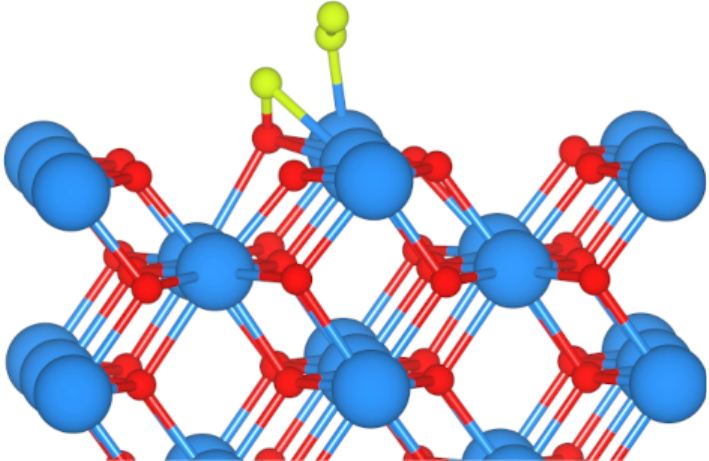}}
        \hspace{0.3 cm}
        \subfloat[c]{\includegraphics[width= 0.32\textwidth]{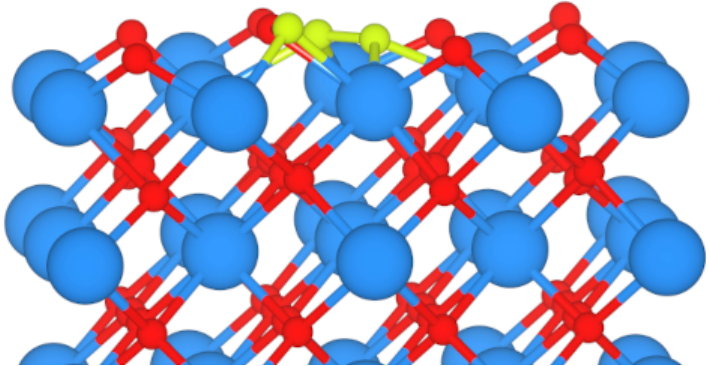}}
        \caption*{Figure S1: Dissociative adsorption of O$_3$ (O$_2$-O) on (a) \{111\}, (b) \{110\}, and (c) \{100\} surfaces of CeO$_2$.}
\end{figure}

\begin{figure}[h!]
\centering
\subfloat[a]{\includegraphics[width= 0.19\textwidth]{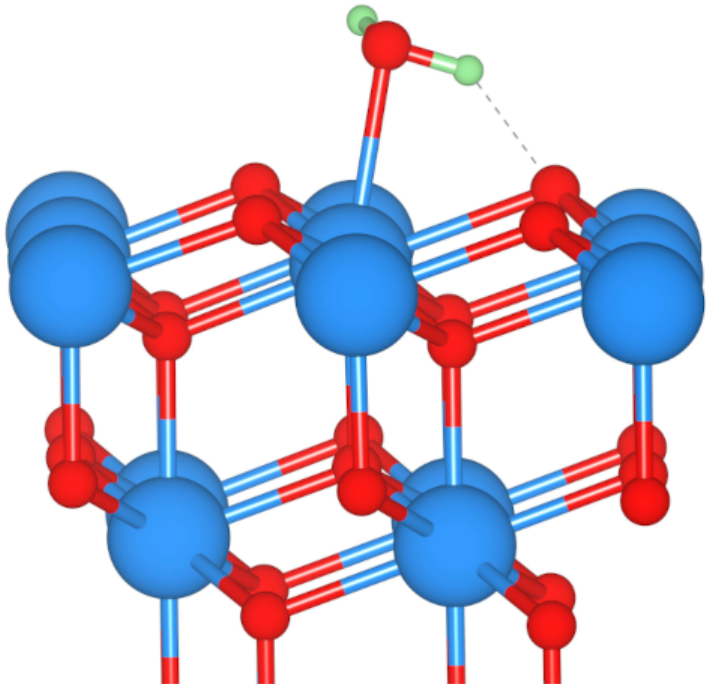}}
        \hspace{0.3 cm}
\subfloat[b]{\includegraphics[width= 0.29\textwidth]{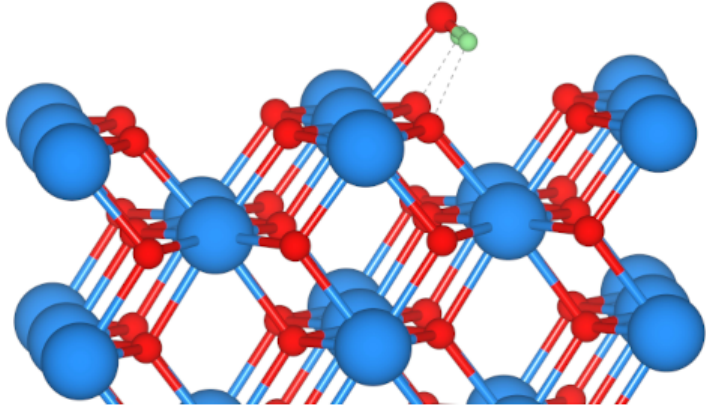}}
        \hspace{0.3 cm}
        \subfloat[c]{\includegraphics[width= 0.31\textwidth]{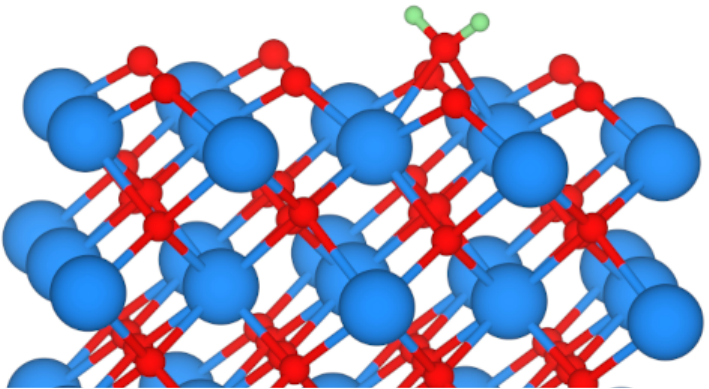}}\\
        \subfloat[d]{\includegraphics[width= 0.22\textwidth]{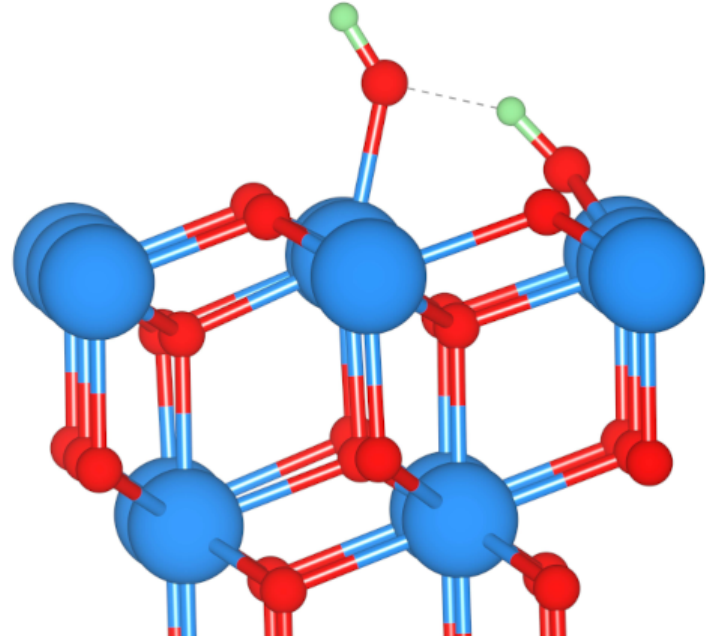}}
        \hspace{0.3 cm}
\subfloat[e]{\includegraphics[width= 0.29\textwidth]{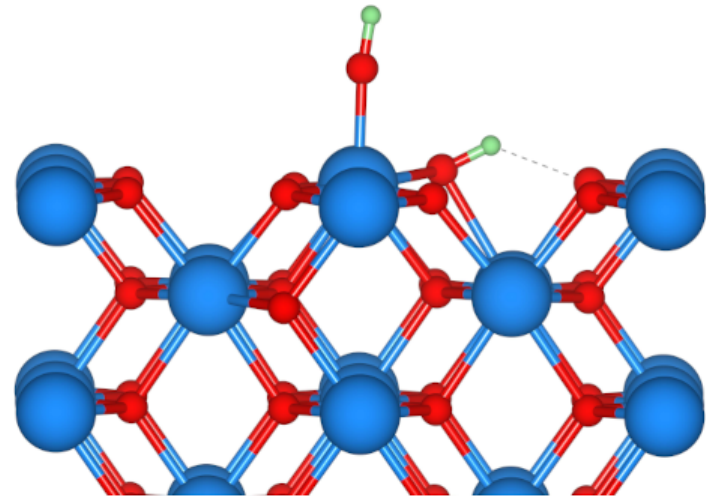}}
        \hspace{0.3 cm}
      \subfloat[f]{\includegraphics[width= 0.31\textwidth]{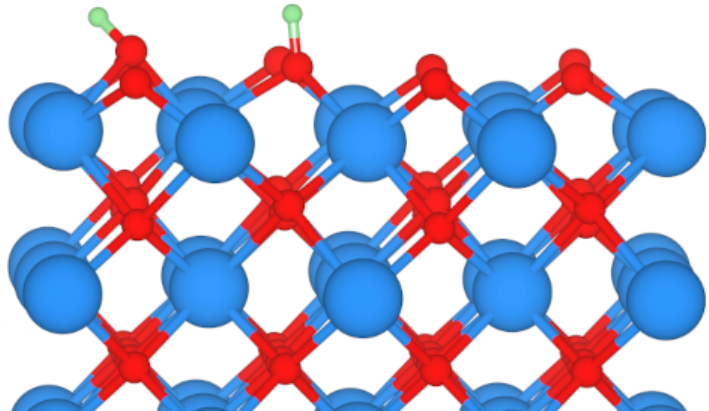}}\\
        \caption*{Figure S2: Stable binding geometries of H$_2$O adsorbed in associate form on (a) \{111\}, (b) \{110\}, and (c) \{100\} surfaces and in dissociative form on (d) \{111\}, (e) \{110\}, and (f) \{100\} surfaces of CeO$_2$.
Hydrogen atoms are represented by green spheres.}
\end{figure}

\begin{figure}[h!]
\centering
\subfloat[a]{\includegraphics[width= 0.19\textwidth]{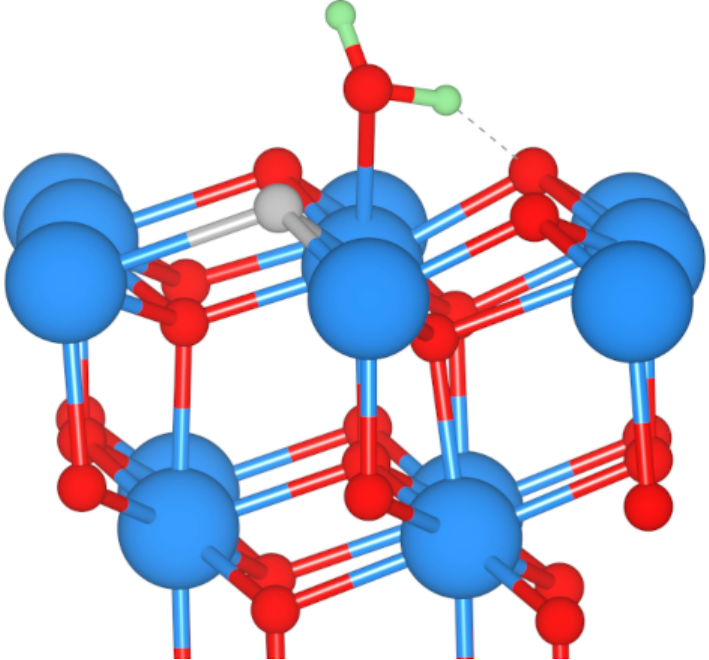}}
        \hspace{0.3 cm}
\subfloat[b]{\includegraphics[width= 0.29\textwidth]{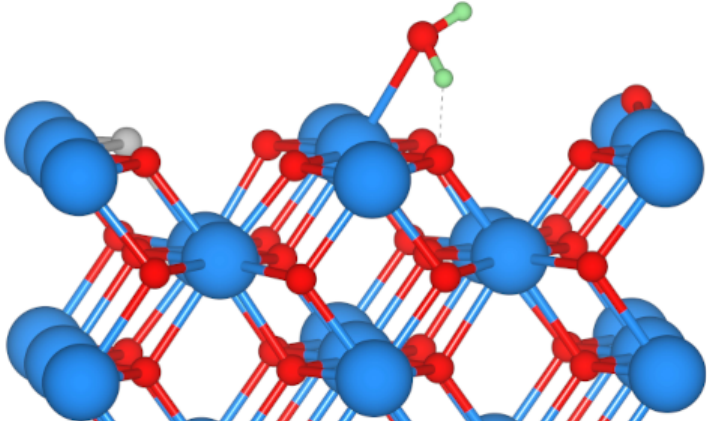}}
        \hspace{0.3 cm}
        \subfloat[c]{\includegraphics[width= 0.31\textwidth]{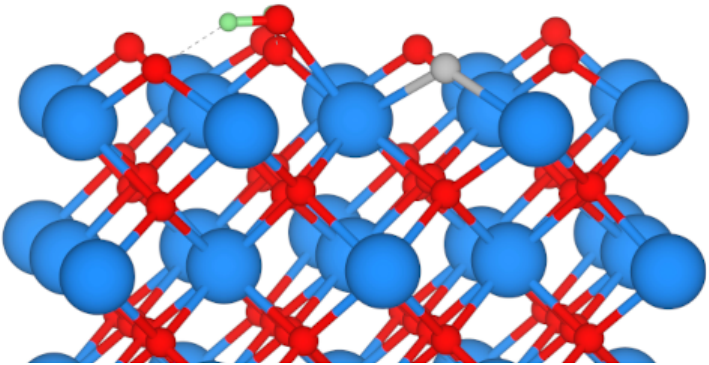}}\\
        \subfloat[d]{\includegraphics[width= 0.22\textwidth]{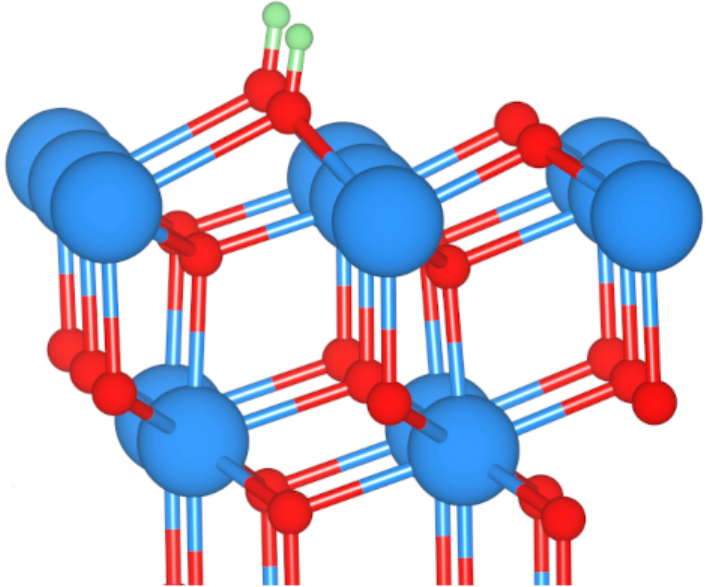}}
        \hspace{0.3 cm}
\subfloat[e]{\includegraphics[width= 0.29\textwidth]{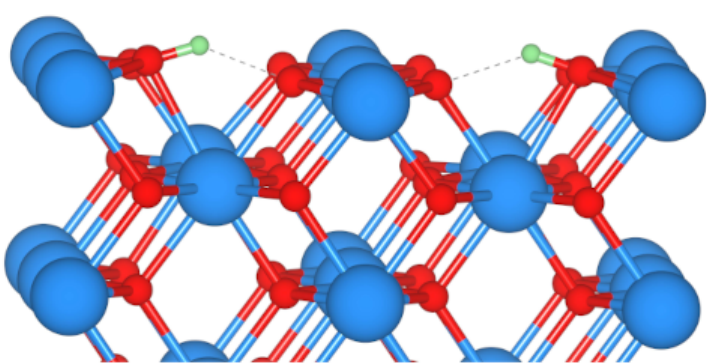}}
        \hspace{0.3 cm}
     \subfloat[f]{\includegraphics[width= 0.31\textwidth]{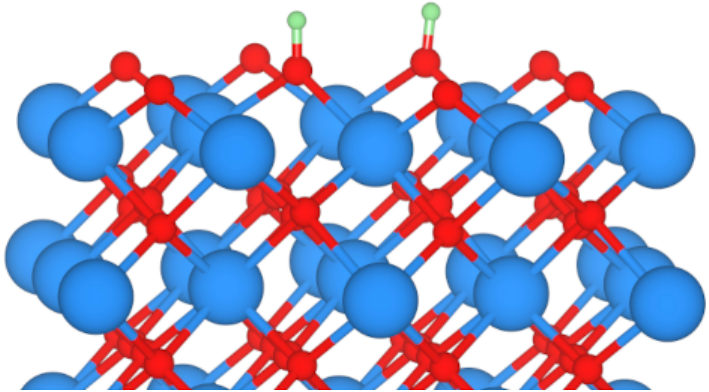}}\\
        \caption*{Figure S3: Stable binding geometries of H$_2$O on (a) \{111\}, (b) \{110\}, and (c) \{100\} surfaces and H-OH on (d) \{111\}, (e) \{110\}, and (f) \{100\} surfaces of reduced CeO$_2$. Oxygen deficient site is represented by grey spheres.}
\end{figure}